\documentclass[aps,prd,twocolumn,amsmath,amssymb,floatfix,nofootinbib,superscriptaddress]{revtex4-1}

\usepackage{graphicx}
\usepackage{dcolumn}
\usepackage{bm}
\usepackage[usenames, dvipsnames]{color}
\usepackage[normalem]{ulem}
\usepackage[dvipsnames]{xcolor}
\usepackage[utf8]{inputenc}
\usepackage{hyperref}
\usepackage{aas_macros}

\newcommand{\be}{\begin{align}}

\newcommand{\ee}{\end{align}}

\newcommand*\dd{\mathop{}\!\mathrm{d}}

\newcommand{\smu}{Department of Physics,
Southern Methodist University, 3215 Daniel Ave, Dallas, TX 75275, USA}

\begin{document}

\title{Searching for Gravitational Waves with Strongly Lensed Repeating Fast Radio Bursts}
\author{Noah~Pearson}
\affiliation{\smu}
\author{Cynthia~Trendafilova}
\affiliation{\smu}
\author{Joel~Meyers}
\affiliation{\smu}

\date{\today}

\begin{abstract}
Since their serendipitous discovery, Fast Radio Bursts (FRBs) have garnered a great deal of attention from both observers and theorists.  A new class of radio telescopes with wide fields of view have enabled a rapid accumulation of FRB observations, confirming that FRBs originate from cosmological distances. The high occurrence rate of FRBs and the development of new instruments to observe them create opportunities for FRBs to be utilized for a host of astrophysical and cosmological studies.  We focus on the rare, and as yet undetected, subset of FRBs that undergo repeated bursts and are strongly gravitationally lensed by intervening structure.  An extremely precise timing of burst arrival times is possible for strongly lensed repeating FRBs, and we show how this timing precision enables the search for long wavelength gravitational waves, including those sourced by supermassive black hole binary systems.  The timing of burst arrival for strongly lensed repeating FRBs is sensitive to gravitational wave sources near the FRB host galaxy, which may lie at cosmological distances and would therefore be extremely challenging to detect by other means.  Timing of strongly lensed FRBs can also be combined with pulsar timing array data to search for correlated time delays characteristic of gravitational waves passing through the Earth.
\end{abstract}

\maketitle

\section{Introduction}

Fast Radios Bursts (FRBs) are a recently discovered class of astrophysical phenomena consisting of radio-wave pulses of roughly millisecond duration, similar in some respects to those from pulsars~\cite{Petroff:2019tty}. Unlike pulsars, however, their large dispersion measures suggest most FRBs are sourced far outside our own Galaxy and thus have much higher luminosity than pulsars. The astrophysical origin of FRBs is currently uncertain~\cite{Platts:2018hiy}, though there is some evidence that there is a connection with magnetars; see \cite{Bochenek:2020zxn}, for example.  Astronomers have discovered more than 100 FRBs, and observations from several radio telescopes are being used to add to the rapidly growing catalog~\cite{Petroff:2016tcr}\footnote{\label{frbcat}\url{http://www.frbcat.org/}}. The list of confirmed FRBs includes a sub-population of FRB sources which have been observed to emit repeated bursts~\cite{Spitler:2016dmz,Amiri:2019bjk,Andersen:2019yex}. 

Many of the current known FRBs have been discovered by existing telescopes that are utilized to search for pulsars and are suitable for detecting FRBs as well, including the Parkes telescope~\cite{Thornton:2013iua}, Arecibo Observatory~\cite{Spitler:2014fla}, and the Green Bank Telescope~\cite{Masui:2015kmb}. However, a new generation of radio telescopes with much wider fields of view, such as the Canadian Hydrogen Intensity Mapping Experiment (CHIME)~\cite{Amiri:2018qsq}, have begun to detect FRBs at a much higher rate and will rapidly increase the known population of sources.

CHIME observations have already been very useful for finding FRBs, with a detection rate of $2-42$~FRBs~$\mathrm{sky}^{-1} \mathrm{day}^{-1}$~\cite{Amiri:2018qsq}.  The majority of known repeating FRBs have been found in CHIME data~\cite{Andersen:2019yex}.  Planned surveys with large fields of view are predicted to be similarly useful or even better at finding FRBs. For example, the Hydrogen Intensity and Real-time Analysis eXperiment (HIRAX) could potentially be used to discover dozens of new FRBs per day, and it will also be capable of measuring the pulse arrival times and their spatial distribution~\cite{Newburgh:2016mwi}.  The Tianlai experiment is similarly suited to finding FRBs~\cite{2012IJMPS..12..256C}.  The Square Kilometre Array (SKA)~\cite{2009IEEEP..97.1482D} with an extremely wide field of view of several hundred square-degrees (at $1000$~MHz) is also expected to be very efficient at finding FRBs~\cite{2016arXiv161000683T,Hashimoto:2020dud}. SKA will cover a lower range of frequencies ($\sim1000$~MHz) compared to most instruments which have been used to find FRBs to date. Most FRBs have been found at 1400~MHz, with the exception of FRB110523 discovered by the Green Bank Telescope at 800~MHz~\cite{Masui:2015kmb}. A future experiment like PUMA could detect thousands of FRBs each day~\cite{Ansari:2018ury,Bandura:2019uvb}.

These upcoming searches will provide an expanded population of observed FRBs, which will be very useful in pursuing the open question of the mechanism generating the bursts.  It is worth considering all the additional ways in which this class of phenomena can be further utilized, for both astrophysical and cosmological applications~\cite{Keane:2018jqo}. For example, the conditions in FRB host galaxies and in the intervening medium may impact the observed pulses, thereby revealing properties of distant galaxies~\cite{Madhavacheril:2019buy} and of the intergalactic medium~\cite{Qiang:2020vta}. Repeating FRBs are likely to be particularly useful, since repeated bursts allow for improved sky localization and can be subjected to detailed follow-up studies~\cite{Andersen:2019yex}. With an increase in the number of observed FRBs, there is also the possibility that some observed FRBs will be strongly lensed, resulting in multiple observed images from which we can extract additional information~\cite{Dai:2017twh,Li:2017mek,Zitrin:2018let,Wagner:2018fvv,Liu:2019jka,Oguri:2019fix,Wucknitz:2020spz}.

We focus in this work on multiply-imaged repeating FRBs.  While no such systems have yet been found, a detection in the future is likely to happen.  Continued and expanded FRB searches, development of future surveys well-matched to FRB detection, and increased attention from the astrophysical community are all likely to contribute to many more FRB discoveries in the coming years.  The likelihood that some of these systems will be strongly lensed repeaters is bolstered by the apparent isotropic distribution of detected FRBs~\cite{Newburgh:2016mwi}, magnification bias of lensed systems~\cite{Dai:2017twh}, and significant fraction of FRBs that have been observed to repeat~\cite{Petroff:2016tcr}.  
Recent estimates for the FRB all-sky rate give a range of $29-3200~\mathrm{sky}^{-1} \mathrm{day}^{-1}$~\cite{Petroff:2019tty}.  The probability that a randomly chosen galaxy on the sky is strongly lensed lies in the range $10^{-4}-10^{-3}$~\cite{Liu:2019jka}.  These estimates suggest that several lensed repeaters may be observed within the coming decades.

Several applications of strongly lensed repeating FRBs have been proposed~\cite{Oguri:2019fix}.  These include measuring the motion of the FRB source~\cite{Dai:2017twh}, measuring the Hubble constant~\cite{Li:2017mek}, constraining dark energy~\cite{Liu:2019jka}, and measuring the matter density of the Universe~\cite{Wucknitz:2020spz}.

In this paper we describe another application of strongly lensed repeating FRBs: the search for long-wavelength gravitational waves.  It is expected that we will be able to measure the time delay between the arrival of images of strongly lensed FRBs with extreme precision~\cite{Wucknitz:2020spz}.  Monitoring a strongly lensed repeating FRB offers the possibility to observe how the time delay changes from one burst to the next, thus allowing for the measurement of even tiny perturbations to the time delay.  Strongly lensed repeating FRBs thereby act as extremely precise clocks that reside at cosmological distances.  Just as the time of arrival of pulses from pulsars in our own Galaxy can be used to search for gravitational waves, we can utilize the time delay between the arrival of images of strongly lensed FRBs to search for gravitational waves that pass through the Milky Way or the FRB host galaxy.

In Section~\ref{sec:PTAs}, we review how pulsar timing can be used to search for gravitational waves, which provides a useful analogue for the case of interest in this paper.  In Section~\ref{sec:Ideal}, we provide details of how monitoring strongly lensed repeating FRBs could be used to search for gravitational waves.  We compare to the use of Galactic pulsars and show that there are gravitational wave sources to which FRBs are much more sensitive.  We discuss some of the practical issues which could complicate the use of strongly lensed repeating FRBs for gravitational wave searches in Section~\ref{sec:TimeDelay}.  We conclude in Section~\ref{sec:Conclusion}.

\section{Pulsars and Gravitational Wave Searches}\label{sec:PTAs}

The use of pulsar timing to search for gravitational waves provides a close analogy to our proposal to use strongly lensed repeating FRBs for the same purpose, and so we briefly review how pulsar observations may be used to detect gravitational waves.  Radio pulsars are rapidly rotating neutron stars that emit beamed radiation which is observed on Earth as a regular series of pulses~\cite{Lorimer:2012book}.  Observed pulse arrival times are compared to models for the emission and propagation of the pulses, and the difference between the observations and the model provides a set of timing residuals.   Millisecond pulsars are a particularly stable subset of pulsars whose very regular rotational period allows for very precise timing measurements over long periods of repeated observation~\cite{Lorimer:2008se}.

The influence of gravitational waves on the propagation of electromagnetic radiation can advance or delay the arrival time of pulsed radiation compared to expectations, thereby generating timing residuals in pulsar timing data~\cite{1975GReGr...6..439E,1978SvA....22...36S,1979ApJ...234.1100D}.  For a single pulsar, it is impossible to uniquely determine the cause of an observed timing residual.  

However, regular measurements of multiple pulsars allow one to search for correlated timing residuals which cannot result from the conditions unique to individual pulsars.  Gravitational waves passing through the Earth are expected to produce a particular pattern of correlated timing residuals~\cite{Hellings:1983fr}.  This motivates the simultaneous and regular monitoring of pulse arrival times from many pulsars across the sky -- a Pulsar Timing Array~\cite{Lommen:2015gbz,Hobbs:2017zve,Tiburzi:2018txc}.  The International Pulsar Timing Array~\cite{Hobbs_2010}, which consists of the Parkes Pulsar Timing Array~\cite{Reardon:2015kba}, the European Pulsar Timing Array~\cite{Desvignes:2016yex}, and the North American Nanohertz Observatory for Gravitational Waves (NANOGrav)~\cite{Arzoumanian:2018saf}, comprises timing data for 65 millisecond pulsars observed for periods ranging from about 2 to more than 20 years.  Based on the observing cadence and span of observations, pulsar timing arrays are sensitive to gravitational waves in the nHz to $\mu$Hz frequency range.  

Sources of gravitational waves can be broadly divided into continuous sources, burst sources, and stochastic backgrounds.  Gravitational wave sources of cosmological origin include inflation, cosmological phase transitions, and cosmic strings~\cite{Maggiore:1999vm}. One of the main targets for pulsar timing arrays is the detection of long wavelength gravitational waves from supermassive black hole binary systems (see~\cite{Burke-Spolaor:2018bvk} for a review). These binaries form from major mergers of galaxies that each contain a supermassive black hole at their center. The gravitational waves from supermassive binary black holes can produce two kinds of signals: a stochastic background, which results from the superposition of a large number of unresolved binaries, and a continuous signal from a single binary that may be close enough or massive enough to be resolved above the background.

The timing residual expected in a pulsar timing array due to a massive black hole binary gravitational wave source is given (up to geometric factors) by
\begin{equation}\label{eq:delay_estimate}
    \Delta \tau \sim 10 \,\mathrm{ns} 
    \left( \frac{1 \,\mathrm{Gpc}}{{d_\mathrm{BBH}}} \right) 
    \left( \frac{M}{10^9 \,M_{\odot}} \right)^{5/3} 
    \left( \frac{10^{-7} \,\mathrm{Hz}}{f} \right)^{1/3} \, ,
\end{equation}
where $d_\mathrm{BBH}$ is the luminosity distance to the binary, $M$ is the total mass of the binary, and $f$ is the observed frequency of the gravitational wave radiation~\cite{2009arXiv0909.1058J}. 

Pulsar timing array data has been used to place a 95\% upper limit of $1.0\times10^{-15}$ on the strain amplitude of a stochastic background of gravitational waves at a frequency of 1~yr$^{-1}\simeq32$~nHz~\cite{Shannon:2015ect,Lentati:2015qwp,Arzoumanian:2015liz,Verbiest:2016vem}.   Analysis of more recent data from NANOGrav shows evidence for a common-spectrum process with equivalent strain amplitude of $1.92\times10^{-15}$ at 32~nHz, though there is not yet sufficient evidence of quadrupolar spatial correlation to conclude that this represents a detection of a stochastic gravitational wave background~\cite{Arzoumanian:2020vkk}.   Estimates of supermassive binary black hole merger rates and the evolution of pulsar timing data suggest that a first detection of long wavelength gravitational waves using pulsar timing arrays will likely occur within the next few years~\cite{Taylor:2015msb}.

\section{Strongly Lensed FRBs and Gravitational Waves}\label{sec:Ideal}

If an FRB is strongly lensed, we will observe multiple images of the same burst on Earth. In the absence of gravitational waves and other complicating factors such as dispersion, the time delay between images is determined by the lensing geometry. This time delay can be measured very precisely~\cite{Wucknitz:2020spz}, and repeated measurements allow us to detect even small variations to this quantity; any change over time may indicate the passing of gravitational waves. For this reason, our method requires measurements of FRBs which are strongly lensed, so that we may use the strong lensing time delay as a precise clock. Measuring the time delay corresponding to only a single burst, however, only gives information about a single point in time and therefore cannot be used to search for gravitational waves. Thus, we require the strongly lensed FRB to be repeating as well; then, we hope to precisely measure the time delay between images for subsequent bursts and look for evolution of the observed time delay. Multiple repetitions would allow for more measurements of the time delay with which to look for any variation. An observation of such a change could indicate the passing of a long-wavelength gravitational wave, while the absence of such a detection can be used to set limits on the stochastic gravitational wave background.

We do not require that the images of the strongly lensed FRB be spatially resolved, nor do we require an accurate lens model in order for the proposed method to be useful.  Furthermore, our method does not rely on regularity of bursts from repeating FRBs, which is important since observed repeaters do not appear to be particularly regular~\cite{Spitler:2016dmz,Scholz:2016rpt,Andersen:2019yex,Fonseca:2020cdd}, despite the periodic activity of one FRB observed by CHIME~\cite{Amiri:2020gno}.  Once the lensing time delay is empirically determined from measurements of one or more lensed bursts, observations of subsequent bursts can be used to search for variations to the time delay.  Of course, gravitational waves are only one potential source of a variation in the time delay.  We focus in this section on the time delay due to gravitational waves, and we return to some complicating factors and their mitigation in Sec.~\ref{sec:TimeDelay}.

\begin{figure}[t!]
    \centering
    \includegraphics[scale=.5]{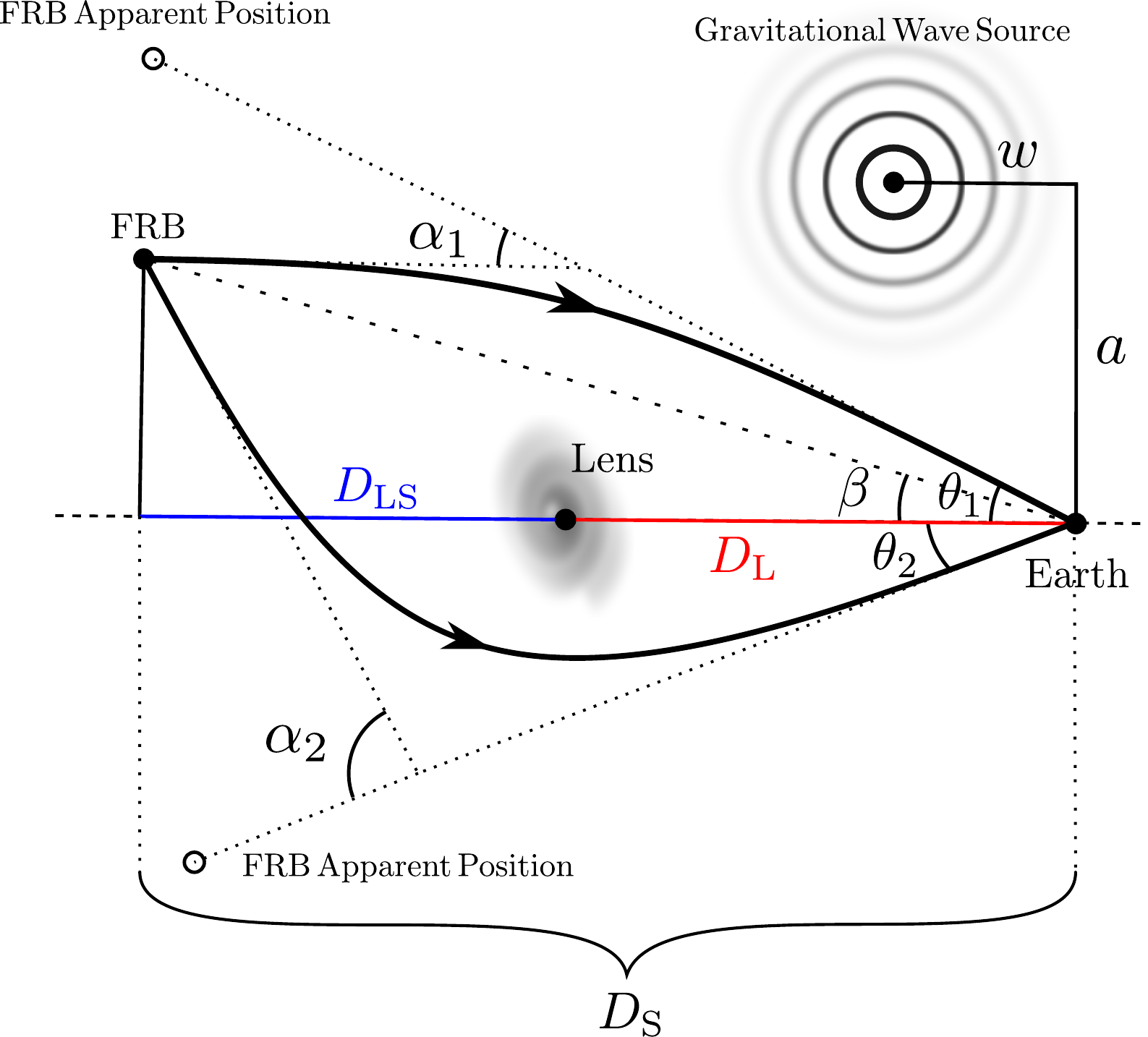}
    \caption{Sketch of the geometry of a strongly lensed FRB and gravitational wave source.  The (unobservable) angle between the lens and the true FRB position is denoted by $\beta$, the angles between the lens and the apparent position of the images of the FRB are $\theta_1$ and $\theta_2$, and the deflection angles near the lens are $\alpha_1$ and $\alpha_2$.  The angular diameter distances from Earth to the lens and source are $D_\mathrm{L}$ and $D_\mathrm{S}$, respectively, and the angular diameter distance between the lens and source is $D_\mathrm{LS}$.  The coordinates of the gravitational wave source parallel to and perpendicular to the line of sight to the lens are $w$ and $a$, respectively.}
    \label{fig:2paths}
\end{figure}

To this end, we consider the geometry illustrated in Figure~\ref{fig:2paths}. The FRB source sits at some cosmological distance away from Earth, indicated by $D_{\mathrm{S}}$. The burst is strongly lensed by an object (such as a galaxy) located at a distance $D_{\mathrm{L}}$ away from Earth, and a gravitational wave source (such as a binary black hole system) is located at a distance $a$ perpendicular to the line of sight to the lens, and a distance $w$ parallel to the line of sight. The lensing results in two or more images of the FRB, observed on Earth at different times and coming from nearby apparent positions in the sky. At some later time the FRB may repeat, and we once again observe multiple images, but with different time delays between them, with the magnitude of the time delays being determined by the gravitational wave amplitude and geometric factors.  

A simplified geometry is presented and compared to the analogous situation using pulsar timing in Figure~\ref{fig:fixed}. The quantities $w$ and $a$ still indicate the position of the gravitational wave source relative to Earth, while the distance from Earth to the FRB is given by $D$. To include a pulsar timing array, we consider an idealized case where we have enough pulsars, isotropically distributed about the Earth, so that we can always choose one whose line of sight relative to us is perpendicular to the line of sight from Earth to the gravitational wave source. This would give the maximal effect to the timing residuals of this pulsar and therefore provides the most optimistic numerical value for comparing with our FRB method.

In this section, we present the calculations necessary to arrive at the change in the time delay over subsequent bursts, due to the influence of gravitational waves. We provide numerical results using a continuous gravitational wave source of constant frequency, and we compare this to the performance of pulsar timing given the same source.

\subsection{Strong gravitational lensing}\label{stronglensing}
Gravitational lensing is the deflection of light by inhomogeneities in gravitational potential. Whenever light is deflected by a sufficiently large mass (such as near a galaxy or galaxy cluster) and multiple images are produced as a result, this is referred to as strong gravitational lensing.

We can make several approximations in describing strong gravitational lensing. In the thin lens approximation, one assumes that the thickness of the lens is very small in comparison to the distances involved between source, lens, and observer. Furthermore, if the deflection angle is small, one can also apply the weak-field approximation for the gravitational field.

We define two-dimensional vectors $\boldsymbol{\beta}$, indicating the position on the sky of the source as it would be seen if there were no lensing (which is unobservable); $\boldsymbol{\theta}$, indicating the position of the image; and $\boldsymbol{\alpha}$, denoting the deflection angle experienced by a light ray near the lens as seen by the lens. These quantities are related by the lens equation,
\begin{equation}
    \boldsymbol{\beta} = \boldsymbol{\theta} - \nabla \psi = \boldsymbol{\theta} - \boldsymbol{\alpha} \, ,
\end{equation}
where $\psi$ is the two-dimensional lensing potential. It satisfies the two-dimensional Poisson equation $\nabla^2 \psi = 2 \kappa$, where $\kappa$ is the surface projected mass density of the deflecting lens, normalized to the critical density $\Sigma_c = c^2 D_\mathrm{S} / (4 \pi G D_\mathrm{L} D_\mathrm{LS})$, where $D_\mathrm{L}$ and $D_\mathrm{S}$ are the angular diameter distances to the lens and the source, respectively and $D_\mathrm{LS}$ is the angular diameter distance between the lens and the source~\cite{2010ARA&A..48...87T}. When multiple values of $\boldsymbol{\theta}$ satisfy the lens equation, multiple images are produced, with common configurations including two images, four images, and a partial or complete Einstein ring. Galaxies and galaxy clusters are sufficiently massive that when they act as gravitational lenses, the images can be separated by more than an arcsecond and are thus individually resolvable~\cite{Oguri:2019fix}.

Since multiple images travel along different paths to the observer, they also have different arrival times. The time delay of an image, compared to the unperturbed path it would take in the absence of lensing, is given by
\begin{equation}
    \Delta t = \frac{(1+z_\mathrm{L})}{c}\frac{D_\mathrm{L} D_\mathrm{S} }{ D_\mathrm{LS}} \left(\frac{1}{2} |\boldsymbol{\theta} - \boldsymbol{\beta}|^2 - \psi(\boldsymbol{\theta}) \right),
\end{equation}
where $z_\mathrm{L}$ is the redshift of the lens~\cite{2010ARA&A..48...87T}. This time delay on its own is not observable; however, we can observe the time delay between pairs of images, as $\Delta t_{i j} = \Delta t_i - \Delta t_j$, with $i$ and $j$ denoting two images. Note that in order to observe this difference in time delay, some manner of time-variation is required from the source. Otherwise, if the signal from the source is constant in time, we cannot find any discernible time delay between its images. For strong lensing by an intervening galaxy, we expect typical time delays ranging from several days to months~\cite{Dai:2017twh}.

\subsection{Timing Precision}

As will be discussed in the next section, the time delay caused by gravitational waves is expected to be on the order of microseconds or smaller.  In order that strongly lensed repeating FRBs may be used to search for gravitational waves, it must be possible to very precisely determine the arrival time of the images.

Since bursts from observed FRBs have a duration on the order of milliseconds, it should be expected that burst arrival times can be measured with uncertainty that is no worse than about a millisecond.  This is impressive precision, but it would require an exceptionally loud gravitational wave source to cause a time delay of this magnitude.  It is fortunate that even better measurements of the time delay between images of a strongly lensed FRB are possible.

More precise time delay measurements are enabled by the fact that gravitationally lensed images of a common source are mutually coherent~\cite{1985A&A...148..369S}.  Such coherence allows for the possibility of utilizing the information contained in the full electromagnetic waves of the arriving images, rather than just the intensity.  Measuring the group delay between the wave images leads to a timing uncertainty which scales as the inverse bandwidth of the observations, while using the phase delay could in principle allow for uncertainties as small as the inverse of the signal frequency of the burst~\cite{Wucknitz:2020spz}. 

According to Ref.~\cite{Wucknitz:2020spz}, a conservative estimate of the precision with which we can measure the time delay between the arrival of two images of a strongly lensed FRB is $10^{-6}~$s, with uncertainties smaller than a nanosecond being achievable. Achieving this precision requires measurements of strongly lensed FRBs with high signal-to-noise ratios, and there are a number of complicating factors which need to be very precisely accounted for, some of which are discussed in Section~\ref{sec:TimeDelay}.  Nevertheless, the possibility for such exquisite timing precision makes strongly lensed FRBs powerful tools for a range of cosmological and astrophysical applications~\cite{Dai:2017twh,Li:2017mek,Zitrin:2018let,Wagner:2018fvv,Liu:2019jka,Oguri:2019fix,Wucknitz:2020spz} including the search for gravitational waves.

\subsection{Gravitational Wave Time Delay}

We now calculate the time delay between images of a strongly lensed system due to gravitational waves.  In general, the geometry is complicated since we want to calculate the effect of gravitational waves on at least two distinct geodesics which may not be co-planar and which are not necessarily short compared to the curvature of the gravitational wave fronts.  However, as long as the gravitational wave is a small perturbation to the metric, the time delay splits into two terms which account for the impact of the gravitational wave on the two endpoints of the photon's path; these terms are usually called the `pulsar term' and the `Earth term' of the gravitational wave time delay when applied to pulsar timing searches for gravitational waves~\cite{1979ApJ...234.1100D,Anholm:2008wy}.  This splitting allows for a simplification to the treatment of the geometry.  

\begin{figure}[t!]
    \centering
    \includegraphics[scale=.6]{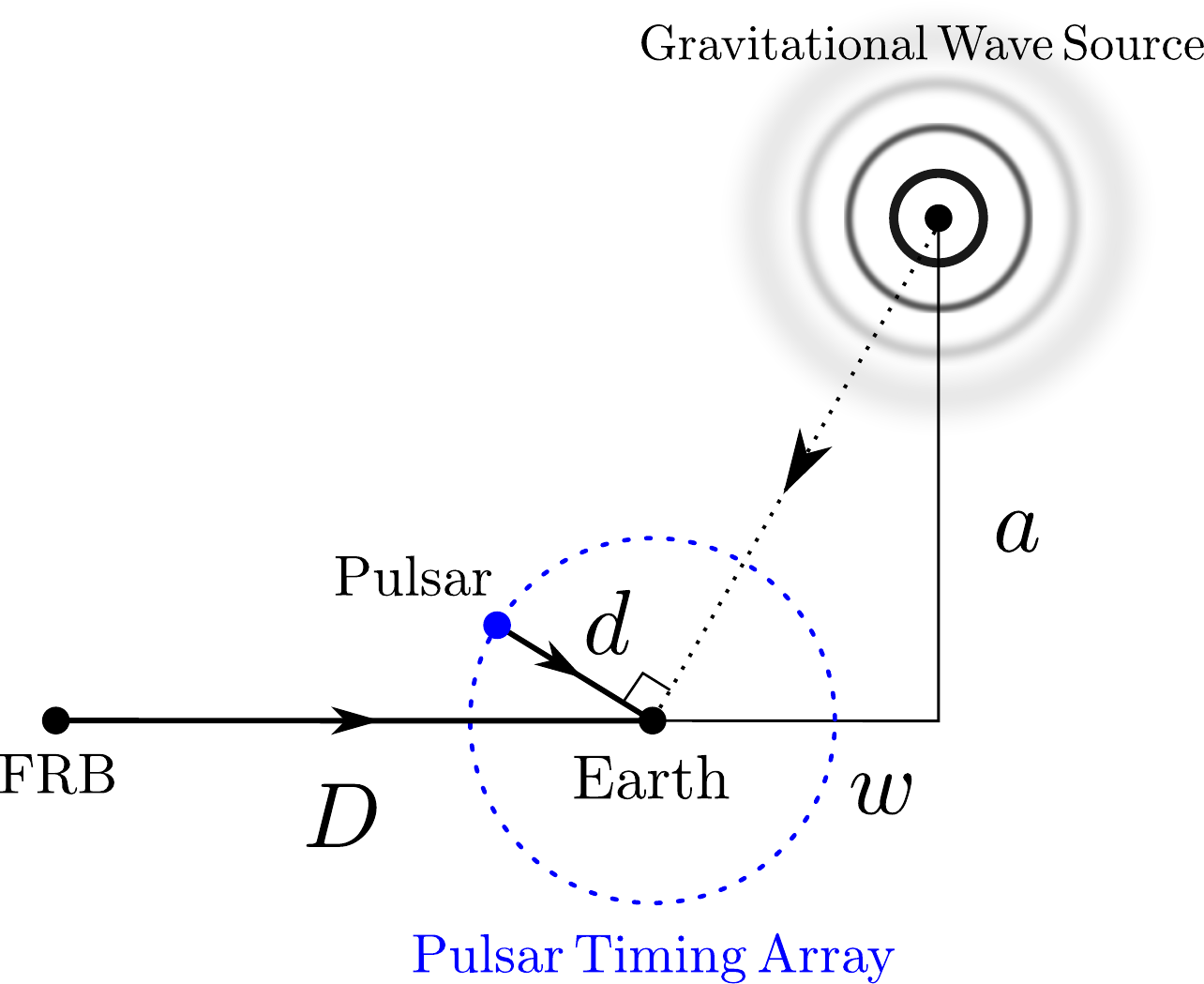}
    \caption{Schematic representation of the geometry used for gravitational wave time delay calculations.  We fix the distance to the FRB $D$ to 100~Mpc and the distance to typical pulsars in a pulsar timing array $d$ to 1~kpc.  The gravitational wave source is taken to be at a point with spatial coordinates $w$ parallel to the line of sight to the FRB and $a$ perpendicular to the FRB line of sight.  We consider an idealized pulsar timing array such that for every gravitational wave source position, there exists a pulsar in a direction perpendicular to the line of sight to the gravitational wave source. 
    }
    \label{fig:fixed}
\end{figure}

Rather than calculating the effect of the gravitational wave on two geodesics of the FRB photons, we instead calculate the gravitational time delay by treating the two strongly lensed images as if they were two subsequent bursts from a single unlensed source.  This treatment is easily justified for the contribution from the Earth term, where the two images of the FRB are well separated in time and arrive with an angular separation on the order of an arcsecond.  The `pulsar term' requires a bit more consideration for strongly lensed FRBs.  When measuring timing residuals with pulsars, one measures the time delay between separate pulses, meaning that the pulses at the source are well separated in time and therefore subject to different phases of the incoming gravitational wave.  For strongly lensed FRBs, it is the same burst that produces both images, and therefore the photons that make up each image are subject to the same gravitational wave at the source.  However, as these geodesics of the images separate, they are subject to independent phases of the gravitational wave.  For example, for arcsecond separation of strongly lensed images, the geodesics of the images will be separated by more than the wavelength of a nHz gravitational wave after traveling about a Mpc from the source.  We therefore expect a contribution to the gravitational wave time delay due to something very much like the ordinary pulsar term, though it will arise due to the differing effects of the gravitational wave on the two photon paths nearby, rather than directly at, the source of the strongly lensed images.  We have verified this heuristic explanation by explicit calculation on a set of simple lensing geometries and have found good agreement with the subsequent pulse approximation\footnote{We also found numerical evidence for a third contribution to the gravitational wave timing delay in the strong lensing geometry which could be called a `lens term' arising from the effect of the gravitational wave near the lens.  This may have resulted from our simplified treatment of the geometry as a set of straight segments, or due to our simple accounting of Shapiro delay which was added by hand at the closest approach to the lens.  We do not include this lens term in our analysis, and the main conclusions are not affected by gravitational wave effects near the lens.}.  

We now briefly describe the calculation of the gravitational wave time delay following that of Ref.~\cite{1978SvA....22...36S}.  In the presence of gravitational waves, described by a perturbation to the metric $h(t,x)$, null geodesics (i.e.~photon paths) are defined as
\begin{equation}
    c\,\dd t=\left[1+\frac{1}{2}h(t,x)\right]\dd x \, .
\end{equation}
Fixing the coordinate positions of the source and observer, the difference in travel time for two subsequent light pulses in the presence of gravitational waves is then
\begin{equation}
    c\,\dd (\delta t)=\frac{1}{2}[h(t_1,x)-h(t_2,x)]\dd x \, .
    \label{eq:cddt}
\end{equation}

Following Ref.~\cite{1978SvA....22...36S}, we model a binary black hole with constant orbital angular frequency $\omega$, such that the gravitational waves produced have angular frequency $2\omega$. The gravitational waves in the plane of the binary are then
\begin{align}
  h(t,x) & =-\frac{\alpha}{2}\frac{a^4-a^2x^2}{(a^2+x^2)^{5/2}}\cos\left[2\omega\left(t-\frac{\sqrt{x^2+a^2}}{c}\right)\right] \, , 
\end{align}
where $a$ is the distance between the binary and the line of sight to the FRB, and 
\begin{align}
    %\alpha & = \frac{3\cdot2^{10/3}\pi^{4/3}\omega^{2/3}G^{5/3}\mathcal{M}^{5/3}}{c^{4}} \, ,
    \alpha & = \frac{12(G\mathcal{M})^{5/3}\omega^{2/3}}{c^{4}} \, ,
\end{align}
with $\mathcal{M}\equiv\frac{(M_1M_2)^{3/5}}{(M_1+M_2)^{1/5}}$ the chirp mass of the binary.

Taking the unperturbed separation in the arrival times of the two images to be $\Delta$ and integrating Eq.~\eqref{eq:cddt} along the path from the FRB to Earth, we find the time delay due to gravitational waves to be
\begin{align}
    \delta(t) =& \frac{\alpha}{2c}\sin{(\omega\Delta)}\int_{-D-w}^{-w} \dd x \, \frac{a^4-a^2x^2}{(a^2+x^2)^{5/2}} \nonumber \\ 
     &\times\sin{\left[2\omega t+2\frac{\omega}{c}(x-\sqrt{x^2+a^2})\right]} \, .
    \label{tdel}
\end{align}
As to be expected, the time delay oscillates with the observation time.  We calculate the root mean square of the time delay in order to give an estimate of its amplitude independent of the unknown phase.

\subsection{Results}

\begin{figure*}[t!]
    \centering
    \includegraphics[width = \textwidth]{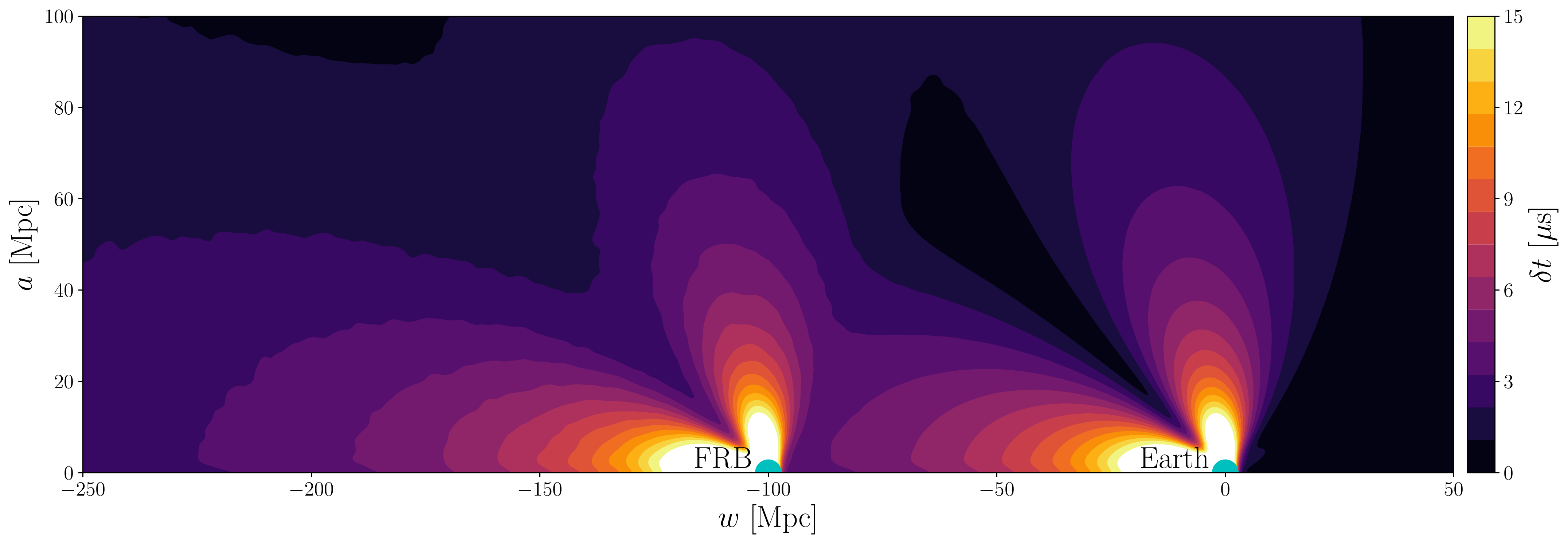}
    \caption{Characteristic time delay between the arrival of the images of a strongly lensed FRB at a distance of 100~Mpc from Earth caused by the gravitational waves from a supermassive binary black hole system.  We plot the root mean square time delay induced by the gravitational waves from a pair of $10^{10}~M_\odot$ black holes with orbital period of 1 year, assuming that the strong lensing time delay between the arrival of FRB images is 0.25 years.  Notice that significant time delay is caused by gravitational wave sources both near the Earth and near the FRB.}
    \label{fig:time_delay}
\end{figure*}

\begin{figure*}[t!]
    \centering
    \includegraphics[width = \textwidth]{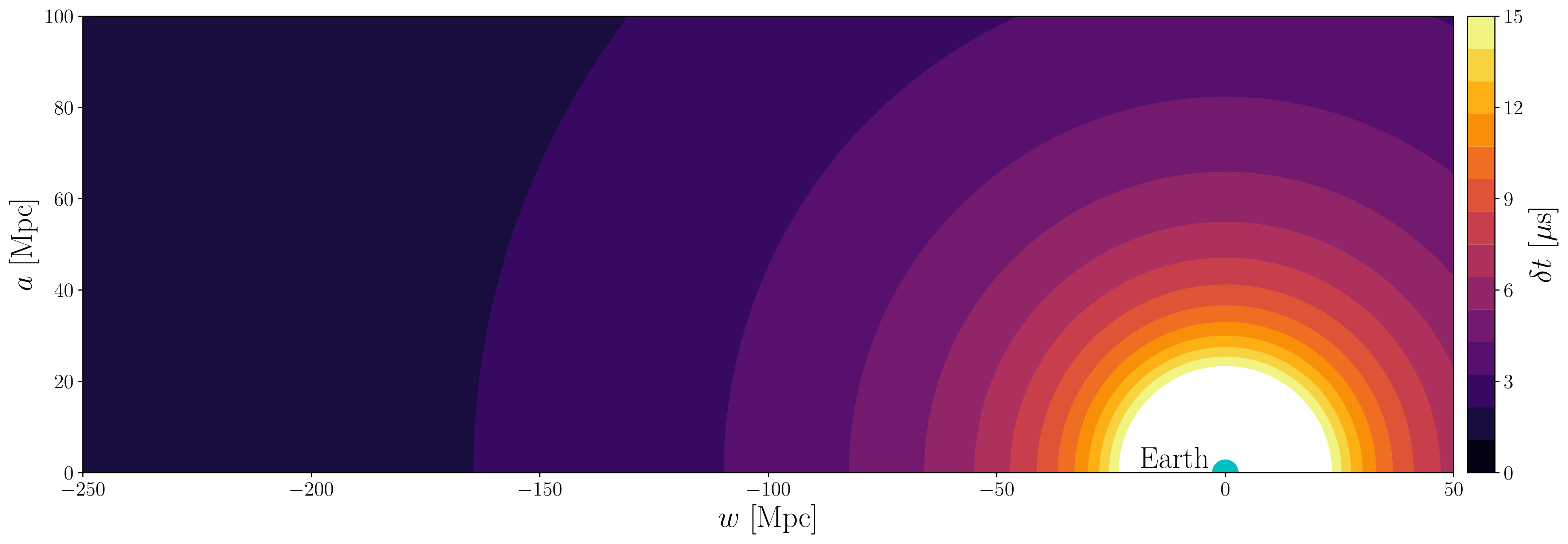}
    \caption{Maximum timing residual for an isotropic pulsar timing array with a typical Earth-pulsar distance of 1~kpc for pulses separated by 0.25 years induced by gravitational waves from a supermassive binary black hole system, using the same assumptions about the gravitational wave source as in Figure~\ref{fig:time_delay}.  Naturally, pulsar timing arrays are most sensitive to gravitational wave sources near the Milky Way.}
    \label{fig:time_delay_pulsar}
\end{figure*}

We now present the time delay between images of a strongly lensed FRB expected due to gravitational waves, and we compare those results to the timing residuals expected in a pulsar timing array.  A sketch of the geometry we consider is shown in Figure~\ref{fig:fixed}.  We fix the distance to the FRB to be $D=100$~Mpc, though there is nothing special about this distance, and qualitative conclusions would be unchanged for more distant FRBs. We consider an idealized pulsar timing array, always assuming that there is a pulsar along a line of sight perpendicular to the line of sight to the gravitational wave source. We take the distance to the pulsars in the pulsar timing array to be $d=1$~kpc, which is typical for currently monitored pulsars~\cite{Corbin:2010kt,Lee:2011et,Matthews:2015kba}.  To calculate the time delay expected for pulsars, we replace $a$ with $\sqrt{a^2+w^2}$ and change the limits of integration to $[-d,0]$ in  Eq.~\eqref{tdel}.

We take the unperturbed time between arrival of strongly lensed FRB images and pulsar pulses to be $\Delta=0.25$~yr.  We focus on the case of gravitational waves sourced by a binary system of supermassive black holes with $M_1=M_2=10^{10}~M_{\odot}$ and orbital period of 1~yr.  In Figure~\ref{fig:time_delay} we show the root mean square time delay due to gravitational waves for the images of a strongly lensed FRB for a range of gravitational wave source positions (defined as shown in Figure~\ref{fig:fixed} with $w$ and $a$ the coordinate positions of the gravitational wave source parallel to and perpendicular to the FRB line of sight, respectively).  We show the analogous time delay for an ideally positioned pulsar in Figure~\ref{fig:time_delay_pulsar}; the time delay for pulsars not situated perpendicular to the line of sight to the gravitational wave source would be reduced by geometric factors compared to the values shown in Figure~\ref{fig:time_delay_pulsar}.

One can clearly see the effects of the `pulsar term' and the Earth term in Figure~\ref{fig:time_delay}, each of which provides sensitivity to gravitational wave sources in a different region of space.  The sensitivity to gravitational wave sources near the host galaxies of strongly lensed repeating FRBs is unique since FRBs lie at cosmological distances, and these systems are therefore sensitive to gravitational wave sources that are difficult or impossible to observe by other means.  The time delays generated by such sources would not be expected to correlate with timing residuals of pulsars in the Milky Way.  Gravitational waves sourced near the Earth, however, would be expected to produce time delays which are correlated across pulsar and FRB timing.  One could therefore effectively treat strongly lensed repeating FRBs as part of the data set when analyzing pulsar timing arrays.  Since pulsars and FRBs have a different nature and are subject to a different set of observational systematic uncertainties, it would be a very useful cross-check to observe a potential gravitational wave signal that produces a time delay for both types of systems.

While we have focused here on a single continuous source of gravitational waves, the same principles should apply equally well to burst sources or a stochastic background of gravitational waves.  For example, observations that do not indicate a time delay from gravitational waves could be used to place an upper bound on the stochastic gravitational wave background near Earth in the present epoch and also near the FRB at the time the burst was emitted (which could be a much earlier epoch for high redshift FRBs).

\section{Complications}\label{sec:TimeDelay}

In this section, we discuss some of the potential issues for achieving the precise timing required to utilize strongly lensed FRBs to search for gravitational waves.  We leave aside the Roemer, Einstein, and Shapiro delays which arise also in pulsar timing and can be modeled with sufficient precision so as not to limit the precision of the measurements we discuss~\cite{Stairs:2003eg}.  A fuller discussion of some of the complicating factors presented here as well as some that we do not address can be found in Refs.~\cite{Dai:2017twh,Zitrin:2018let,Wucknitz:2020spz}.

When an electromagnetic pulse propagates through space, components of different frequencies will travel at different speeds due to dispersion experienced by propagation through the interstellar medium (ISM) and to a lesser extent the intergalactic medium.  Dispersion results in pulse broadening due to the fact that different frequencies have different arrival times at Earth. Typically, for a cold, unmagnetized plasma, the time delay scales like the inverse square of the frequency, $t_{\mathrm{DM}}(\nu) = K \nu^{-2} \mathrm{DM}$. Here $\mathrm{DM}$ is the dispersion measure $\mathrm{DM}=\int \frac{n_e(r,\hat{\mathbf{n}})}{1+z} \dd l$ with $n_e$ the number density of electrons along the line of sight~\cite{Inoue:2003ga}, and $K = c\, r_e / 2\pi$, where $r_e$ is the classical electron radius~\cite{Cordes:2015hia}. To determine arrival times as accurately as possible, one must dedisperse pulses to remove this effect.

There are several commonly used methods for attempting to determine and remove the delays imposed on FRBs due to their propagation through the dispersive medium of the ISM. Dedispersion techniques have long been developed and used for dedispersing pulsar signals. Traditionally, dedispersion was performed via algorithms which can be classified as incoherent dedispersion, in contrast to more recently developed methods of coherent dedispersion~\cite{Stairs:2002abcd,Petroff:2019tty}.

With incoherent dedispersion, the frequency bandwidth of the observations is divided into channels, and time delays are removed from these channels on an individual basis. The time delay for each sub-band can be calculated and subtracted, so the time delays between channels get removed, while leaving the dispersive delay within the individual channels. Incoherent dedispersive methods are the preferred choice for pulses whose dispersion measure is not already known.

On the other hand, coherent dedispersion removes the dispersive effect entirely, thus allowing for better timing and precision. While a pulse is propagating through the ISM, the dispersion effect changes its phase in a way that can be modeled by a transfer function, or filter. This transfer function depends on the dispersion measure and characterizes the dispersive time delays observed in the signal. This allows one to take raw voltage data and convolve it with the inverse of this transfer function, thus reversing the effect of the dispersion across all frequencies and recovering the original shape of the pulse with greater precision. Because coherent dedispersion methods can be computationally costly, especially for signals with high dispersion measures (as is true for FRBs), their practicality in blind searches is limited. Instead, they are better suited for cases where the dispersion measure of the burst has already been measured. Repeating FRBs, which are the primary interest of this paper, should have a well-measured dispersion measure that can be used as the starting point for coherent dedispersion of subsequent bursts.

Although we can expect an uncertainty of $10^{-9}~$--$10^{-6}~$s in measuring the time delay between the arrival times of images of a strongly lensed FRB~\cite{Wucknitz:2020spz}, there are many complicating factors which can affect this precision. 

\subsection{Frequency-dependent dispersion measure}

So long as the electron density is spatially uniform, the dispersion measure will be independent of frequency; however, in the ISM this is not the case, and there are density variations over scales ranging from kpc to thousands of km~\cite{1995ApJ...443..209A}. These fluctuations on small scales cause scattering that results in slight variations in DM along different propagation paths, and a propagating pulse will sample the ISM differently via multipath scattering. Such conditions can lead to effects that can be modeled as a frequency dependent dispersion measure~\cite{Cordes:2015hia}, and which can become significant for measurements aiming for high timing precision. Typical dispersion measures for FRBs are on the order of hundreds of $\mathrm{pc\,cm}^{-3}$. The results of Ref.~\cite{Cordes:2015hia} suggest that bursts subject to $\mathrm{DM}\sim300$~$\mathrm{pc\,cm}^{-3}$ and with frequencies of order GHz result in uncertainties on the time of arrival that could be as large as ms or higher without mitigation.  Characterization of the frequency dependence is possible with an additional frequency-dependent term when fitting for arrival times and would allow for the impact on timing measurements to be reduced.

\subsection{Dispersion measure variations over time}
Since the precision of measuring the time delay between strongly lensed FRB images depends upon the coherence of the bursts, one point of concern comes from the possibility that the dispersion measure along the line of sight to the FRB may change with time. If the time delay is of order months, there could be non-negligible evolution of the dispersion measure between the arrival of the two burst images. In Ref.~\cite{Yang_2017}, the authors studied several effects which can cause a time-dependent dispersion measure, including both large-scale and local effects.

Among large-scale effects, the Hubble expansion rate can change the dispersion measure by a rate of order $-5.6 \times 10^{-8}(1+z)^3~\mathrm{pc}\,\mathrm{cm}^{-3}\mathrm{yr}^{-1}$, which for typical FRB redshifts is quite small. Other large-scale effects, such as fluctuations in the gravitational potential of large-scale structures, result in dispersion measure changes which are even smaller than the effect from the Hubble expansion, and therefore negligible.

For the local effects studied, the authors consider an FRB embedded in a supernova remnant, pulsar wind nebula, and local HII region in its host galaxy. A supernova remnant could have an effect as large as $-52~\mathrm{pc}\,\mathrm{cm}^{-3}\,\mathrm{yr}^{-1}$, which may cause a noticeable evolution. An FRB near a relatively young pulsar could have its dispersion measure affected by a pulsar wind nebula by as much as $-11~\mathrm{pc}\,\mathrm{cm}^{-3}\,\mathrm{yr}^{-1}$. Finally, a young HII region surrounding the FRB could have an effect of order $0.78~\mathrm{pc}\,\mathrm{cm}^{-3}\,\mathrm{yr}^{-1}$.   Effects such as these that are local to the FRB host galaxy should not cause a large change in the observed dispersion measure for images resulting from a common burst, but they may cause an evolution of the dispersion measure for subsequent bursts from the same system.

\subsection{Effects of the lens galaxy}
There are several factors which may contribute to a broadening of the received pulse, including gravitational microlensing by stars located in the galaxy which acts as the gravitational lens and also scattering by the interstellar medium of the lensing galaxy. If the pulse image is located such that at most one star is causing microlensing, the resulting micro-images have a typical time delay of $7 \times 10^{-3}~\mathrm{ms}$~\cite{Dai:2017twh}. On the other hand, if the image propagates close to within the center of the lens galaxy, one may get multiple strongly coupled microlenses and micro-images. These result in a temporal broadening of the FRB which, though not expected to occur very often, can be as large as ms.

Another complication arises from scattering by the ISM of the lens galaxy, which can cause broadening due to the turbulent electron density fluctuations in the ISM. This effect is especially significant at lower frequencies. At frequencies of order 1~GHz, the temporal broadening can be greater than 30~ms, while at 3~GHz, it decreases to about 0.25~ms for the same choice of parameters~\cite{Dai:2017twh}. Particularly at lower frequencies, precision in the timing accuracy could suffer, and it is also possible that some images become too temporally spread out to be detectable. One must also consider scattering of the pulse in the host galaxy, which can cause additional temporal broadening, but this effect will contribute in the same manner to all images resulting from a common burst, and it becomes negligible at frequencies $\gtrsim 3~ \mathrm{GHz}$~\cite{Dai:2017twh}.

\subsection{Plasma lensing}

In addition to the effects of the ISM on dispersion measure, one should also consider the consequences of plasma lensing, caused by scattering from overdensities (or underdensities) of electrons with respect to the mean density of the ISM~\cite{2017ApJ...842...35C}. This lensing can occur as the pulse propagates in both the host galaxy of the FRB and in the Milky Way galaxy, with the effects determined by the density and shape of the plasma lens, and the geometry between source, lens, and observer. Any plasma lensing that occurs in the host galaxy, prior to the FRB getting gravitationally lensed, would not affect the precision of time delay measurements between images of a single burst. However, plasma lensing that occurs as bursts propagate through the ISM of the Milky Way can potentially affect images of the same burst differently, thus limiting how precisely their relative time delay can be determined.

One effect of plasma lensing is the potential to result in multiple subimages of the original pulse, arriving with slightly perturbed arrival times and with slight variation in their observed dispersion measures. In particular, if the time delays between these subimages are very small, one may observe interference between the images. Furthermore, regardless of whether or not the burst is multiply imaged by the plasma lens, it will also experience perturbations to arrival time and dispersion measure which are dependent on frequency, further complicating the processes of coherently dedispersing the bursts and measuring their time delay.

Because plasma lensing is most likely to occur along lines of sight with large Galactic dispersion measure, the optimal circumstance for observing an FRB would be along a line of sight that is perpendicular to the plane of the Milky Way. This provides the greatest likelihood that a gravitationally lensed FRB would not experience significant plasma lensing, allowing the time delay between images of the burst to be determined with greater precision.

\subsection{Relative velocities}

For a repeating, strongly lensed FRB, relative motion between the FRB source, the lens, and the Earth will induce a variation in the time delay between pairs of images~\cite{Dai:2017twh,Zitrin:2018let,Wucknitz:2020spz}. Any motions that are uniform over the observational period of interest will result in linear drifts in the time delay.
A line-of-sight velocity of the source or lens induces a change in the time delay of order $0.01~ \mathrm{ms}$, while a transverse velocity induces a time delay change of order seconds. In addition to this, the motion of the Earth's orbit around the Sun generates a sinusoidal perturbation in the time delay, which can be as large as $10^3~\mathrm{s}$. However, it is shown in Ref.~\cite{Dai:2017twh} that multiple images, resolved with very-long-baseline interferometry, would allow for this effect to be subtracted with ms accuracy.  The rotation of the Earth around its axis also induces a variation in the time delay, which can be as large as $40~\mathrm{ms}$, and would need to be accounted for as well. At worst, these effects of the Earth's motion will limit the sensitivity of strongly lensed FRBs to gravitational waves with periods closely matching the annual and daily motions of the Earth.

There are additional effects, including sources of nonuniform motion and the passing of gravitational waves, which will cause a nontrivial, nonlinear change in the time delay between repetitions. With only two repetitions of a strongly lensed FRB, one cannot determine whether any apparent change in the time delay is linear. Continuous monitoring of a strongly lensed system to observe several bursts is therefore desirable to help distinguish a linear drift from a nonlinear perturbation.

\section{Signal to Noise}
We now present an estimate of the signal-to-noise ratio with which the effects of a continuous gravitational wave source on the arrival times of images of a strongly lensed repeating FRB could be observed. In principle, any pair of images resulting from a common burst can be used to search for the effects of gravitational waves on the observed time delay.  At least one burst must be measured to determine a baseline time delay between the arrival of the images to which subsequent burst timings can be compared.  Furthermore, as discussed above, relative velocities between the source, lens, and observer are expected to induce a drift in the observed time delays, even in the absence of gravitational waves.  We therefore require at least one additional burst to measure linear changes to the time delays, and only subsequent bursts can be used to measure the nonlinear changes to the time delays that are expected from passing gravitational waves.

We will assume that $N_\mathrm{bursts}$ distinct burst events are observed and that we see $n$ images of each burst due to gravitational lensing.  We treat the occurrence of bursts as a stochastic process, such that the arrival times of images from different bursts are random and independent. We can then estimate the signal-to-noise ratio with which we observe the effect of gravitational waves to be
\begin{equation*}
    \left(\frac{S}{N}\right)^2 = (N_\mathrm{bursts}-2)\sum_{i<j\leq n} \left(\frac{\delta(t_{ij})}{\sigma_\Delta}\right)^2 \, ,
\end{equation*}
where $i,j=\{1,\ldots,n\}$ label the individual images resulting from a single burst, $\delta(t_{ij})$ is the perturbation to the time delay between the arrival of images $i$ and $j$ due to the passing of a gravitational wave as given by Eq.~\eqref{tdel}, and $\sigma_\Delta$ is the precision with which the time delay between images of a single burst can be measured.  Eq.~\eqref{eq:delay_estimate} shows the dependence of the expected time delay on the mass of the binary that acts as the source of gravitational waves, the gravitational wave frequency, and the distance between the binary and either the FRB or the Earth.  Figure~\ref{fig:time_delay} shows how the relative position of the binary, the FRB, and the Earth affects the amplitude of the time delay.  

The unperturbed time delay between the arrival of images compared to the frequency of the gravitational wave plays an important role in determining the amplitude of the perturbation to the time delay, and therefore the detectability of gravitational waves.  Specifically, Eq.~\eqref{tdel} contains a factor $\sin(\omega\Delta)$, where $\omega$ is the angular frequency of the gravitational wave and $\Delta$ is the difference in the arrival time between two images of the same burst.  This factor implies that images whose unperturbed arrival times are separated by much less than the period of a gravitational wave do not experience a large relative time delay due to the gravitational wave.  A strongly lensed repeating FRB for which there are several observed images with arrival times separated by intervals comparable to about one half of the gravitational wave period would be ideal, since the amplitude of the time delay would be maximized, and the oscillating behavior of the time delay could be observed.

This simplified treatment of the signal-to-noise ratio is designed to show how the detectability of gravitational waves using strongly lensed repeating FRBs scales with the various quantities that determine the physical situation.  However, as discussed in Section~\ref{sec:TimeDelay}, there are various complicating effects that are more easily mitigated with continued observations.  Long-term monitoring of such systems is likely to be required in order to achieve the timing precision necessary to isolate the small effect of gravitational waves.  

\section{Discussion and Conclusions}\label{sec:Conclusion}

In this work, we have shown the potential of using strongly lensed repeating FRBs to detect long-wavelength gravitational waves. The time delay between images of strongly lensed FRBs can be measured with very high precision, and therefore observations of strongly lensed repeating FRBs offer the possibility of measuring the time evolution of the lensing time delay. This exquisite timing precision enables the search for the effects of long-wavelength gravitational waves, such as those produced by the merger of two supermassive black holes.

The passage of such gravitational waves through the Earth and through the FRB host galaxy causes a change in the observed time delay between burst images. For the parameter space of the illustrative case considered in this paper, the resulting time delay variation can be upwards of 15~$\mu\mathrm{s}$, large enough to be observable given current experimental sensitivity. Unlike pulsars, which are most sensitive to gravitational wave sources near the Earth, strongly lensed repeating FRB systems are sensitive to gravitational waves passing through Earth and through the FRB source, which may reside at cosmological distances. Additionally, observing a strongly lensed repeating FRB and monitoring it over repeated bursts creates the possibility of correlating this timing data with the timing residuals measured in pulsar timing arrays. This could improve the sensitivity to detection of gravitational waves near Earth, or it could also further improve constraints derived from non-detection via either method.

We have discussed several factors which may affect the timing precision of these time delay measurements. However, it is still plausible that we may get measurements with sufficient precision to detect the presence of gravitational waves, since it is possible that these effects may be mitigated or avoided. For example, if the line of sight to an FRB is perpendicular to the plane of the Milky Way, this reduces the presence of intervening plasma responsible for plasma lensing. Additionally, the conditions responsible for these timing affects are interesting in and of themselves. The effects of microlensing on the timing measurements, for example, are useful in revealing information about the substructure which is responsible for the lensing; monitoring a repeating FRB over time and identifying the effect of source velocity would allow one to follow its orbit and perhaps learn something about its environment and origins~\cite{Dai:2017twh}.

We described how strongly lensed repeating FRBs can be used to search for signals from long-wavelength gravitational waves. For this purpose, a change in the time delay could indicate a potential detection. However, the time delay caused by the passing of gravitational waves would be a source of noise in other proposed applications of strongly lensed repeating FRBs where very high timing precision is desired. Detecting gravitational waves via the method proposed here or with pulsar timing array measurements could perhaps help characterize the effective noise from gravitational waves that could hinder other applications and experiments.

Certain effects pertaining to the propagation of gravitational waves, whose potential detection has been considered in the case of pulsar timing array observations, would also affect the gravitational waves that can be detected by observing strongly lensed repeating FRBs. These include the effects of certain cosmological parameters on gravitational wave propagation, such as the cosmological constant $\Lambda$, the density of non-relativistic matter $\Omega_c$, and the expansion history more generally~\cite{Alfaro:2017sxd,Alfaro:2019sbq}. The effects of these parameters on timing residuals allow for local measurements of the relevant cosmological parameters, and the high timing precision possible with FRB measurements could be leveraged, in tandem with pulsar timing array measurements, to improve the detection prospects of these effects.

FRBs have already proven to be an exciting new target for the astrophysical community. The situation is likely to improve significantly in the coming years as current and planned experiments will vastly expand the catalog of observed FRBs, a catalog which may contain several FRBs which are both repeating and strongly lensed.  Strongly lensed repeating FRBs would be valuable for a range of astrophysical and cosmological applications~\cite{Dai:2017twh,Li:2017mek,Zitrin:2018let,Wagner:2018fvv,Liu:2019jka,Oguri:2019fix,Wucknitz:2020spz}. The search for gravitational waves described in this work provides additional motivation to seek out and monitor these systems to take advantage of the unique opportunities they offer.

\section*{Acknowledgments}
The authors would like to thank Simon Foreman, Selim Hotinli, and Ue-Li Pen for helpful conversations.  The authors are also grateful to Evan Grohs and Marios Kalomenopoulos for valuable comments and corrections on earlier versions.  This work is supported by the US Department of Energy under grant no.~DE-SC0010129 and by the Hamilton Undergraduate Research Scholars Program.

\bibliographystyle{utphys}
\bibliography{refs}

\providecommand{\href}[2]{#2}\begingroup\raggedright\begin{thebibliography}{10}

\bibitem{Petroff:2019tty}
E.~Petroff, J.~W.~T. Hessels, and D.~R. Lorimer, ``{Fast Radio Bursts},''
  \href{http://dx.doi.org/10.1007/s00159-019-0116-6}{{\em Astron. Astrophys.
  Rev.} {\bfseries 27} no.~1, (2019) 4},
\href{http://arxiv.org/abs/1904.07947}{{\ttfamily arXiv:1904.07947
  [astro-ph.HE]}}.
%%CITATION = ARXIV:1904.07947;%%.

\bibitem{Platts:2018hiy}
E.~Platts, A.~Weltman, A.~Walters, S.~Tendulkar, J.~Gordin, and S.~Kandhai,
  ``{A Living Theory Catalogue for Fast Radio Bursts},''
  \href{http://dx.doi.org/10.1016/j.physrep.2019.06.003}{{\em Phys. Rept.}
  {\bfseries 821} (2019) 1--27},
  \href{http://arxiv.org/abs/1810.05836}{{\ttfamily arXiv:1810.05836
  [astro-ph.HE]}}.

\bibitem{Bochenek:2020zxn}
C.~D. Bochenek, V.~Ravi, K.~V. Belov, G.~Hallinan, J.~Kocz, S.~R. Kulkarni, and
  D.~L. McKenna, ``{A fast radio burst associated with a Galactic magnetar},''
  \href{http://arxiv.org/abs/2005.10828}{{\ttfamily arXiv:2005.10828
  [astro-ph.HE]}}.

\bibitem{Petroff:2016tcr}
E.~Petroff, E.~Barr, A.~Jameson, E.~Keane, M.~Bailes, M.~Kramer, V.~Morello,
  D.~Tabbara, and W.~van Straten, ``{FRBCAT: The Fast Radio Burst Catalogue},''
  \href{http://dx.doi.org/10.1017/pasa.2016.35}{{\em Publ. Astron. Soc.
  Austral.} {\bfseries 33} (2016) e045},
  \href{http://arxiv.org/abs/1601.03547}{{\ttfamily arXiv:1601.03547
  [astro-ph.HE]}}.

\bibitem{Spitler:2016dmz}
L.~Spitler {\em et~al.}, ``{A Repeating Fast Radio Burst},''
  \href{http://dx.doi.org/10.1038/nature17168}{{\em Nature} {\bfseries 531}
  (2016) 202}, \href{http://arxiv.org/abs/1603.00581}{{\ttfamily
  arXiv:1603.00581 [astro-ph.HE]}}.

\bibitem{Amiri:2019bjk}
{\bfseries CHIME/FRB} Collaboration, M.~Amiri {\em et~al.}, ``{A Second Source
  of Repeating Fast Radio Bursts},''
  \href{http://dx.doi.org/10.1038/s41586-018-0864-x}{{\em Nature} {\bfseries
  566} no.~7743, (2019) 235--238},
  \href{http://arxiv.org/abs/1901.04525}{{\ttfamily arXiv:1901.04525
  [astro-ph.HE]}}.

\bibitem{Andersen:2019yex}
{\bfseries CHIME/FRB} Collaboration, B.~C. Andersen {\em et~al.}, ``{CHIME/FRB
  Detection of Eight New Repeating Fast Radio Burst Sources},''
  \href{http://dx.doi.org/10.3847/2041-8213/ab4a80}{{\em Astrophys. J.}
  {\bfseries 885} no.~1, (2019) L24},
\href{http://arxiv.org/abs/1908.03507}{{\ttfamily arXiv:1908.03507
  [astro-ph.HE]}}.
%%CITATION = ARXIV:1908.03507;%%.

\bibitem{Thornton:2013iua}
D.~Thornton {\em et~al.}, ``{A Population of Fast Radio Bursts at Cosmological
  Distances},'' \href{http://dx.doi.org/10.1126/science.1236789}{{\em Science}
  {\bfseries 341} no.~6141, (2013) 53--56},
  \href{http://arxiv.org/abs/1307.1628}{{\ttfamily arXiv:1307.1628
  [astro-ph.HE]}}.

\bibitem{Spitler:2014fla}
L.~Spitler {\em et~al.}, ``{Fast Radio Burst Discovered in the Arecibo Pulsar
  ALFA Survey},'' \href{http://dx.doi.org/10.1088/0004-637X/790/2/101}{{\em
  Astrophys. J.} {\bfseries 790} no.~2, (2014) 101},
  \href{http://arxiv.org/abs/1404.2934}{{\ttfamily arXiv:1404.2934
  [astro-ph.HE]}}.

\bibitem{Masui:2015kmb}
K.~Masui {\em et~al.}, ``{Dense magnetized plasma associated with a fast radio
  burst},'' \href{http://dx.doi.org/10.1038/nature15769}{{\em Nature}
  {\bfseries 528} (2015) 523},
  \href{http://arxiv.org/abs/1512.00529}{{\ttfamily arXiv:1512.00529
  [astro-ph.HE]}}.

\bibitem{Amiri:2018qsq}
{\bfseries CHIME/FRB} Collaboration, M.~Amiri {\em et~al.}, ``{The CHIME Fast
  Radio Burst Project: System Overview},''
  \href{http://arxiv.org/abs/1803.11235}{{\ttfamily arXiv:1803.11235
  [astro-ph.IM]}}.

\bibitem{Newburgh:2016mwi}
L.~Newburgh {\em et~al.}, ``{HIRAX: A Probe of Dark Energy and Radio
  Transients},'' \href{http://dx.doi.org/10.1117/12.2234286}{{\em Proc. SPIE
  Int. Soc. Opt. Eng.} {\bfseries 9906} (2016) 99065X},
  \href{http://arxiv.org/abs/1607.02059}{{\ttfamily arXiv:1607.02059
  [astro-ph.IM]}}.

\bibitem{2012IJMPS..12..256C}
X.~{Chen}, \href{http://dx.doi.org/10.1142/S2010194512006459}{``{The Tianlai
  Project: a 21CM Cosmology Experiment},''} in {\em International Journal of
  Modern Physics Conference Series}, vol.~12 of {\em International Journal of
  Modern Physics Conference Series}, pp.~256--263.
\newblock Mar., 2012.
\newblock \href{http://arxiv.org/abs/1212.6278}{{\ttfamily arXiv:1212.6278
  [astro-ph.IM]}}.

\bibitem{2009IEEEP..97.1482D}
P.~E. {Dewdney}, P.~J. {Hall}, R.~T. {Schilizzi}, and T.~J.~L.~W. {Lazio},
  ``{The Square Kilometre Array},''
  \href{http://dx.doi.org/10.1109/JPROC.2009.2021005}{{\em IEEE Proceedings}
  {\bfseries 97} no.~8, (Aug., 2009) 1482--1496}.

\bibitem{2016arXiv161000683T}
S.~A. {Torchinsky}, J.~W. {Broderick}, A.~{Gunst}, A.~J. {Faulkner}, and
  W.~{van Cappellen}, ``{SKA Aperture Array Mid Frequency Science
  Requirements},'' {\em arXiv e-prints} (Oct., 2016) arXiv:1610.00683,
  \href{http://arxiv.org/abs/1610.00683}{{\ttfamily arXiv:1610.00683
  [astro-ph.IM]}}.

\bibitem{Hashimoto:2020dud}
T.~Hashimoto, T.~Goto, A.~Y. On, T.-Y. Lu, D.~J.~D. Santos, S.~C.-C. Ho, T.-W.
  Wang, S.~J. Kim, and T.~Y.-Y. Hsiao, ``{Fast radio bursts to be detected with
  the Square Kilometre Array},''
  \href{http://arxiv.org/abs/2008.00007}{{\ttfamily arXiv:2008.00007
  [astro-ph.HE]}}.

\bibitem{Ansari:2018ury}
{R. Ansari \textit{et al.} (Cosmic Visions 21~cm Collaboration)}, ``{Inflation
  and Early Dark Energy with a Stage II Hydrogen Intensity Mapping
  Experiment},''
\href{http://arxiv.org/abs/1810.09572}{{\ttfamily arXiv:1810.09572
  [astro-ph.CO]}}.
%%CITATION = ARXIV:1810.09572;%%.

\bibitem{Bandura:2019uvb}
{\bfseries PUMA} Collaboration, A.~z. Slosar {\em et~al.}, ``{Packed
  Ultra-wideband Mapping Array (PUMA): A Radio Telescope for Cosmology and
  Transients},'' \href{http://arxiv.org/abs/1907.12559}{{\ttfamily
  arXiv:1907.12559 [astro-ph.IM]}}.

\bibitem{Keane:2018jqo}
E.~Keane, ``{The Future of Fast Radio Burst Science},''
  \href{http://arxiv.org/abs/1811.00899}{{\ttfamily arXiv:1811.00899
  [astro-ph.HE]}}.

\bibitem{Madhavacheril:2019buy}
M.~S. Madhavacheril, N.~Battaglia, K.~M. Smith, and J.~L. Sievers, ``{Cosmology
  with the kinematic Sunyaev-Zeldovich effect: Breaking the optical depth
  degeneracy with fast radio bursts},''
  \href{http://dx.doi.org/10.1103/PhysRevD.100.103532}{{\em Phys. Rev. D}
  {\bfseries 100} no.~10, (2019) 103532},
  \href{http://arxiv.org/abs/1901.02418}{{\ttfamily arXiv:1901.02418
  [astro-ph.CO]}}.

\bibitem{Qiang:2020vta}
D.-C. Qiang and H.~Wei, ``{Reconstructing the Fraction of Baryons in the
  Intergalactic Medium with Fast Radio Bursts via Gaussian Processes},''
  \href{http://dx.doi.org/10.1088/1475-7516/2020/04/023}{{\em JCAP} {\bfseries
  04} (2020) 023}, \href{http://arxiv.org/abs/2002.10189}{{\ttfamily
  arXiv:2002.10189 [astro-ph.CO]}}.

\bibitem{Dai:2017twh}
L.~Dai and W.~Lu, ``{Probing motion of fast radio burst sources by timing
  strongly lensed repeaters},''
  \href{http://dx.doi.org/10.3847/1538-4357/aa8873}{{\em Astrophys. J.}
  {\bfseries 847} no.~1, (2017) 19},
\href{http://arxiv.org/abs/1706.06103}{{\ttfamily arXiv:1706.06103
  [astro-ph.HE]}}.
%%CITATION = ARXIV:1706.06103;%%.

\bibitem{Li:2017mek}
Z.-X. Li, H.~Gao, X.-H. Ding, G.-J. Wang, and B.~Zhang, ``{Strongly lensed
  repeating fast radio bursts as precision probes of the universe},''
  \href{http://dx.doi.org/10.1038/s41467-018-06303-0}{{\em Nature Commun.}
  {\bfseries 9} no.~1, (2018) 3833},
\href{http://arxiv.org/abs/1708.06357}{{\ttfamily arXiv:1708.06357
  [astro-ph.CO]}}.
%%CITATION = ARXIV:1708.06357;%%.

\bibitem{Zitrin:2018let}
A.~Zitrin and D.~Eichler, ``{Observing Cosmological Processes in Real Time with
  Repeating Fast Radio Bursts},''
  \href{http://dx.doi.org/10.3847/1538-4357/aad6a2}{{\em Astrophys. J.}
  {\bfseries 866} no.~2, (2018) 101},
  \href{http://arxiv.org/abs/1807.03287}{{\ttfamily arXiv:1807.03287
  [astro-ph.CO]}}.

\bibitem{Wagner:2018fvv}
J.~Wagner, J.~Liesenborgs, and D.~Eichler, ``{Multiply-imaged time-varying
  sources behind galaxy clusters - Comparing FRBs to QSOs, SNe, and GRBs},''
  \href{http://dx.doi.org/10.1051/0004-6361/201833530}{{\em Astron. Astrophys.}
  {\bfseries 621} (2019) A91},
  \href{http://arxiv.org/abs/1811.10618}{{\ttfamily arXiv:1811.10618
  [astro-ph.CO]}}.

\bibitem{Liu:2019jka}
B.~Liu, Z.~Li, H.~Gao, and Z.-H. Zhu, ``{Prospects of strongly lensed repeating
  fast radio bursts: Complementary constraints on dark energy evolution},''
  \href{http://dx.doi.org/10.1103/PhysRevD.99.123517}{{\em Phys. Rev. D}
  {\bfseries 99} no.~12, (2019) 123517},
  \href{http://arxiv.org/abs/1907.10488}{{\ttfamily arXiv:1907.10488
  [astro-ph.CO]}}.

\bibitem{Oguri:2019fix}
M.~Oguri, ``{Strong gravitational lensing of explosive transients},''
  \href{http://dx.doi.org/10.1088/1361-6633/ab4fc5}{{\em Rept. Prog. Phys.}
  {\bfseries 82} no.~12, (2019) 126901},
  \href{http://arxiv.org/abs/1907.06830}{{\ttfamily arXiv:1907.06830
  [astro-ph.CO]}}.

\bibitem{Wucknitz:2020spz}
O.~Wucknitz, L.~Spitler, and U.-L. Pen, ``{Cosmology with gravitationally
  lensed repeating Fast Radio Bursts},''
  \href{http://arxiv.org/abs/2004.11643}{{\ttfamily arXiv:2004.11643
  [astro-ph.CO]}}.

\bibitem{Lorimer:2012book}
D.~Lorimer and M.~Kramer, {\em Handbook of Pulsar Astronomy}.
\newblock Cambridge Observing Handbooks for Research Astronomers. Cambridge
  University Press, 2012.

\bibitem{Lorimer:2008se}
D.~Lorimer, ``{Binary and Millisecond Pulsars},''
  \href{http://dx.doi.org/10.12942/lrr-2008-8}{{\em Living Rev. Rel.}
  {\bfseries 11} (2008) 8}, \href{http://arxiv.org/abs/0811.0762}{{\ttfamily
  arXiv:0811.0762 [astro-ph]}}.

\bibitem{1975GReGr...6..439E}
F.~B. {Estabrook} and H.~D. {Wahlquist}, ``{Response of Doppler spacecraft
  tracking to gravitational radiation.},''
  \href{http://dx.doi.org/10.1007/BF00762449}{{\em General Relativity and
  Gravitation} {\bfseries 6} no.~5, (Oct., 1975) 439--447}.

\bibitem{1978SvA....22...36S}
M.~V. {Sazhin}, ``{Opportunities for detecting ultralong gravitational
  waves},'' {\em Soviet Astronomy} {\bfseries 22} (Feb., 1978) 36--38.

\bibitem{1979ApJ...234.1100D}
S.~{Detweiler}, ``{Pulsar timing measurements and the search for gravitational
  waves},'' \href{http://dx.doi.org/10.1086/157593}{{\em \apj} {\bfseries 234}
  (Dec., 1979) 1100--1104}.

\bibitem{Hellings:1983fr}
R.~Hellings and G.~Downs, ``{Upper limits on the isotropic gravitational
  radiation background from pulsar timing analysis},''
  \href{http://dx.doi.org/10.1086/183954}{{\em Astrophys. J. Lett.} {\bfseries
  265} (1983) L39--L42}.

\bibitem{Lommen:2015gbz}
A.~N. Lommen, ``{Pulsar timing arrays: the promise of gravitational wave
  detection},'' \href{http://dx.doi.org/10.1088/0034-4885/78/12/124901}{{\em
  Rept. Prog. Phys.} {\bfseries 78} no.~12, (2015) 124901}.

\bibitem{Hobbs:2017zve}
G.~Hobbs and S.~Dai, ``{Gravitational wave research using pulsar timing
  arrays},'' \href{http://dx.doi.org/10.1093/nsr/nwx126}{{\em Natl. Sci. Rev.}
  {\bfseries 4} no.~5, (2017) 707--717},
\href{http://arxiv.org/abs/1707.01615}{{\ttfamily arXiv:1707.01615
  [astro-ph.IM]}}.
%%CITATION = ARXIV:1707.01615;%%.

\bibitem{Tiburzi:2018txc}
C.~Tiburzi, ``{Pulsars probe the low-frequency gravitational sky: Pulsar Timing
  Arrays basics and recent results},''
  \href{http://dx.doi.org/10.1017/pasa.2018.7}{{\em Publ. Astron. Soc.
  Austral.} {\bfseries 35} (2018) e013},
\href{http://arxiv.org/abs/1802.05076}{{\ttfamily arXiv:1802.05076
  [astro-ph.IM]}}.
%%CITATION = ARXIV:1802.05076;%%.

\bibitem{Hobbs_2010}
G.~Hobbs, A.~Archibald, Z.~Arzoumanian, D.~Backer, M.~Bailes, N.~D.~R. Bhat,
  M.~Burgay, S.~Burke-Spolaor, D.~Champion, I.~Cognard, and et~al., ``The
  international pulsar timing array project: using pulsars as a gravitational
  wave detector,'' \href{http://dx.doi.org/10.1088/0264-9381/27/8/084013}{{\em
  Classical and Quantum Gravity} {\bfseries 27} no.~8, (Apr, 2010) 084013}.
  \url{http://dx.doi.org/10.1088/0264-9381/27/8/084013}.

\bibitem{Reardon:2015kba}
D.~Reardon {\em et~al.}, ``{Timing analysis for 20 millisecond pulsars in the
  Parkes Pulsar Timing Array},''
  \href{http://dx.doi.org/10.1093/mnras/stv2395}{{\em Mon. Not. Roy. Astron.
  Soc.} {\bfseries 455} no.~2, (2016) 1751--1769},
  \href{http://arxiv.org/abs/1510.04434}{{\ttfamily arXiv:1510.04434
  [astro-ph.HE]}}.

\bibitem{Desvignes:2016yex}
G.~Desvignes {\em et~al.}, ``{High-precision timing of 42 millisecond pulsars
  with the European Pulsar Timing Array},''
  \href{http://dx.doi.org/10.1093/mnras/stw483}{{\em Mon. Not. Roy. Astron.
  Soc.} {\bfseries 458} no.~3, (2016) 3341--3380},
  \href{http://arxiv.org/abs/1602.08511}{{\ttfamily arXiv:1602.08511
  [astro-ph.HE]}}.

\bibitem{Arzoumanian:2018saf}
{\bfseries NANOGRAV} Collaboration, Z.~Arzoumanian {\em et~al.}, ``{The
  NANOGrav 11-year Data Set: Pulsar-timing Constraints On The Stochastic
  Gravitational-wave Background},''
  \href{http://dx.doi.org/10.3847/1538-4357/aabd3b}{{\em Astrophys. J.}
  {\bfseries 859} no.~1, (2018) 47},
  \href{http://arxiv.org/abs/1801.02617}{{\ttfamily arXiv:1801.02617
  [astro-ph.HE]}}.

\bibitem{Maggiore:1999vm}
M.~Maggiore, ``{Gravitational wave experiments and early universe cosmology},''
  \href{http://dx.doi.org/10.1016/S0370-1573(99)00102-7}{{\em Phys. Rept.}
  {\bfseries 331} (2000) 283--367},
  \href{http://arxiv.org/abs/gr-qc/9909001}{{\ttfamily arXiv:gr-qc/9909001}}.

\bibitem{Burke-Spolaor:2018bvk}
S.~Burke-Spolaor {\em et~al.}, ``{The Astrophysics of Nanohertz Gravitational
  Waves},'' \href{http://dx.doi.org/10.1007/s00159-019-0115-7}{{\em Astron.
  Astrophys. Rev.} {\bfseries 27} no.~1, (2019) 5},
\href{http://arxiv.org/abs/1811.08826}{{\ttfamily arXiv:1811.08826
  [astro-ph.HE]}}.
%%CITATION = ARXIV:1811.08826;%%.

\bibitem{2009arXiv0909.1058J}
F.~{Jenet}, L.~S. {Finn}, J.~{Lazio}, A.~{Lommen}, M.~{McLaughlin},
  I.~{Stairs}, D.~{Stinebring}, J.~{Verbiest}, A.~{Archibald},
  Z.~{Arzoumanian}, D.~{Backer}, J.~{Cordes}, P.~{Demorest}, R.~{Ferdman},
  P.~{Freire}, M.~{Gonzalez}, V.~{Kaspi}, V.~{Kondratiev}, D.~{Lorimer},
  R.~{Lynch}, D.~{Nice}, S.~{Ransom}, R.~{Shannon}, and X.~{Siemens}, ``{The
  North American Nanohertz Observatory for Gravitational Waves},'' {\em arXiv
  e-prints} (Sep, 2009) arXiv:0909.1058,
  \href{http://arxiv.org/abs/0909.1058}{{\ttfamily arXiv:0909.1058
  [astro-ph.IM]}}.

\bibitem{Shannon:2015ect}
R.~Shannon {\em et~al.}, ``{Gravitational waves from binary supermassive black
  holes missing in pulsar observations},''
  \href{http://dx.doi.org/10.1126/science.aab1910}{{\em Science} {\bfseries
  349} no.~6255, (2015) 1522--1525},
  \href{http://arxiv.org/abs/1509.07320}{{\ttfamily arXiv:1509.07320
  [astro-ph.CO]}}.

\bibitem{Lentati:2015qwp}
L.~Lentati {\em et~al.}, ``{European Pulsar Timing Array Limits On An Isotropic
  Stochastic Gravitational-Wave Background},''
  \href{http://dx.doi.org/10.1093/mnras/stv1538}{{\em Mon. Not. Roy. Astron.
  Soc.} {\bfseries 453} no.~3, (2015) 2576--2598},
  \href{http://arxiv.org/abs/1504.03692}{{\ttfamily arXiv:1504.03692
  [astro-ph.CO]}}.

\bibitem{Arzoumanian:2015liz}
{\bfseries NANOGrav} Collaboration, Z.~Arzoumanian {\em et~al.}, ``{The
  NANOGrav Nine-year Data Set: Limits on the Isotropic Stochastic Gravitational
  Wave Background},'' \href{http://dx.doi.org/10.3847/0004-637X/821/1/13}{{\em
  Astrophys. J.} {\bfseries 821} no.~1, (2016) 13},
  \href{http://arxiv.org/abs/1508.03024}{{\ttfamily arXiv:1508.03024
  [astro-ph.GA]}}.

\bibitem{Verbiest:2016vem}
J.~Verbiest {\em et~al.}, ``{The International Pulsar Timing Array: First Data
  Release},'' \href{http://dx.doi.org/10.1093/mnras/stw347}{{\em Mon. Not. Roy.
  Astron. Soc.} {\bfseries 458} no.~2, (2016) 1267--1288},
  \href{http://arxiv.org/abs/1602.03640}{{\ttfamily arXiv:1602.03640
  [astro-ph.IM]}}.

\bibitem{Arzoumanian:2020vkk}
{\bfseries NANOGrav} Collaboration, Z.~Arzoumanian {\em et~al.}, ``{The
  NANOGrav 12.5-year Data Set: Search For An Isotropic Stochastic
  Gravitational-Wave Background},''
  \href{http://arxiv.org/abs/2009.04496}{{\ttfamily arXiv:2009.04496
  [astro-ph.HE]}}.

\bibitem{Taylor:2015msb}
S.~Taylor, M.~Vallisneri, J.~Ellis, C.~Mingarelli, T.~Lazio, and R.~van
  Haasteren, ``{Are we there yet? Time to detection of nanohertz gravitational
  waves based on pulsar-timing array limits},''
  \href{http://dx.doi.org/10.3847/2041-8205/819/1/L6}{{\em Astrophys. J. Lett.}
  {\bfseries 819} no.~1, (2016) L6},
  \href{http://arxiv.org/abs/1511.05564}{{\ttfamily arXiv:1511.05564
  [astro-ph.IM]}}.

\bibitem{Scholz:2016rpt}
P.~Scholz {\em et~al.}, ``{The repeating Fast Radio Burst FRB 121102:
  Multi-wavelength observations and additional bursts},''
  \href{http://dx.doi.org/10.3847/1538-4357/833/2/177}{{\em Astrophys. J.}
  {\bfseries 833} no.~2, (2016) 177},
\href{http://arxiv.org/abs/1603.08880}{{\ttfamily arXiv:1603.08880
  [astro-ph.HE]}}.
%%CITATION = ARXIV:1603.08880;%%.

\bibitem{Fonseca:2020cdd}
E.~Fonseca {\em et~al.}, ``{Nine New Repeating Fast Radio Burst Sources from
  CHIME/FRB},''
\href{http://arxiv.org/abs/2001.03595}{{\ttfamily arXiv:2001.03595
  [astro-ph.HE]}}.
%%CITATION = ARXIV:2001.03595;%%.

\bibitem{Amiri:2020gno}
{\bfseries CHIME/FRB} Collaboration, M.~Amiri {\em et~al.}, ``{Periodic
  activity from a fast radio burst source},''
  \href{http://dx.doi.org/10.1038/s41586-020-2398-2}{{\em \nat} {\bfseries 582}
  no.~7812, (June, 2020) 351--355},
  \href{http://arxiv.org/abs/2001.10275}{{\ttfamily arXiv:2001.10275
  [astro-ph.HE]}}.

\bibitem{2010ARA&A..48...87T}
T.~{Treu}, ``{Strong Lensing by Galaxies},''
  \href{http://dx.doi.org/10.1146/annurev-astro-081309-130924}{{\em Annual
  Review of Astronomy and Astrophysics} {\bfseries 48} (Sept., 2010) 87--125},
  \href{http://arxiv.org/abs/1003.5567}{{\ttfamily arXiv:1003.5567
  [astro-ph.CO]}}.

\bibitem{1985A&A...148..369S}
P.~{Schneider} and J.~{Schmid-Burgk}, ``{Mutual coherence of gravitationally
  lensed images},'' {\em \aap} {\bfseries 148} no.~2, (July, 1985) 369--378.

\bibitem{Anholm:2008wy}
M.~Anholm, S.~Ballmer, J.~D. Creighton, L.~R. Price, and X.~Siemens, ``{Optimal
  strategies for gravitational wave stochastic background searches in pulsar
  timing data},'' \href{http://dx.doi.org/10.1103/PhysRevD.79.084030}{{\em
  Phys. Rev. D} {\bfseries 79} (2009) 084030},
  \href{http://arxiv.org/abs/0809.0701}{{\ttfamily arXiv:0809.0701 [gr-qc]}}.

\bibitem{Corbin:2010kt}
V.~Corbin and N.~J. Cornish, ``{Pulsar Timing Array Observations of Massive
  Black Hole Binaries},'' \href{http://arxiv.org/abs/1008.1782}{{\ttfamily
  arXiv:1008.1782 [astro-ph.HE]}}.

\bibitem{Lee:2011et}
K.~Lee, N.~Wex, M.~Kramer, B.~Stappers, C.~Bassa, G.~Janssen, R.~Karuppusamy,
  and R.~Smits, ``{Gravitational wave astronomy of single sources with a pulsar
  timing array},''
  \href{http://dx.doi.org/10.1111/j.1365-2966.2011.18622.x}{{\em Mon. Not. Roy.
  Astron. Soc.} {\bfseries 414} (2011) 3251},
  \href{http://arxiv.org/abs/1103.0115}{{\ttfamily arXiv:1103.0115
  [astro-ph.HE]}}.

\bibitem{Matthews:2015kba}
{\bfseries NANOGrav} Collaboration, A.~M. Matthews {\em et~al.}, ``{The
  NANOGrav Nine-year Data Set: Astrometric Measurements of 37 Millisecond
  Pulsars},'' \href{http://dx.doi.org/10.3847/0004-637X/818/1/92}{{\em
  Astrophys. J.} {\bfseries 818} no.~1, (2016) 92},
  \href{http://arxiv.org/abs/1509.08982}{{\ttfamily arXiv:1509.08982
  [astro-ph.GA]}}.

\bibitem{Stairs:2003eg}
I.~H. Stairs, ``{Testing general relativity with pulsar timing},''
  \href{http://dx.doi.org/10.12942/lrr-2003-5}{{\em Living Rev. Rel.}
  {\bfseries 6} (2003) 5},
  \href{http://arxiv.org/abs/astro-ph/0307536}{{\ttfamily
  arXiv:astro-ph/0307536}}.

\bibitem{Inoue:2003ga}
S.~Inoue, ``{Probing the cosmic reionization history and local environment of
  gamma-ray bursts through radio dispersion},''
  \href{http://dx.doi.org/10.1111/j.1365-2966.2004.07359.x}{{\em Mon. Not. Roy.
  Astron. Soc.} {\bfseries 348} (2004) 999},
  \href{http://arxiv.org/abs/astro-ph/0309364}{{\ttfamily
  arXiv:astro-ph/0309364}}.

\bibitem{Cordes:2015hia}
J.~Cordes, R.~Shannon, and D.~Stinebring, ``{Frequency-Dependent Dispersion
  Measures and Implications for Pulsar Timing},''
  \href{http://dx.doi.org/10.3847/0004-637X/817/1/16}{{\em Astrophys. J.}
  {\bfseries 817} no.~1, (2016) 16},
  \href{http://arxiv.org/abs/1503.08491}{{\ttfamily arXiv:1503.08491
  [astro-ph.IM]}}.

\bibitem{Stairs:2002abcd}
I.~H. {Stairs}, ``{Pulsar Observations II. -- Coherent Dedispersion
  Polarimetry, and Timing},'' in {\em Single-Dish Radio Astronomy: Techniques
  and Applications}, S.~{Stanimirovic}, D.~{Altschuler}, P.~{Goldsmith}, and
  C.~{Salter}, eds., vol.~278 of {\em Astronomical Society of the Pacific
  Conference Series}, pp.~251--269.
\newblock Dec., 2002.

\bibitem{1995ApJ...443..209A}
J.~W. {Armstrong}, B.~J. {Rickett}, and S.~R. {Spangler}, ``{Electron Density
  Power Spectrum in the Local Interstellar Medium},''
  \href{http://dx.doi.org/10.1086/175515}{{\em ApJ} {\bfseries 443} (Apr.,
  1995) 209}.

\bibitem{Yang_2017}
Y.-P. Yang and B.~Zhang, ``Dispersion measure variation of repeating fast radio
  burst sources,'' \href{http://dx.doi.org/10.3847/1538-4357/aa8721}{{\em The
  Astrophysical Journal} {\bfseries 847} no.~1, (Sep, 2017) 22}.
  \url{https://doi.org/10.3847%2F1538-4357%2Faa8721}.

\bibitem{2017ApJ...842...35C}
J.~M. {Cordes}, I.~{Wasserman}, J.~W.~T. {Hessels}, T.~J.~W. {Lazio},
  S.~{Chatterjee}, and R.~S. {Wharton}, ``{Lensing of Fast Radio Bursts by
  Plasma Structures in Host Galaxies},''
  \href{http://dx.doi.org/10.3847/1538-4357/aa74da}{{\em ApJ} {\bfseries 842}
  no.~1, (June, 2017) 35}, \href{http://arxiv.org/abs/1703.06580}{{\ttfamily
  arXiv:1703.06580 [astro-ph.HE]}}.

\bibitem{Alfaro:2017sxd}
J.~Alfaro, D.~Espriu, and L.~Gabbanelli, ``{On the propagation of gravitational
  waves in a $\Lambda$CDM universe},''
  \href{http://dx.doi.org/10.1088/1361-6382/aaf675}{{\em Class. Quant. Grav.}
  {\bfseries 36} no.~2, (2019) 025006},
  \href{http://arxiv.org/abs/1711.08315}{{\ttfamily arXiv:1711.08315
  [hep-th]}}.

\bibitem{Alfaro:2019sbq}
J.~Alfaro and M.~Gamonal, ``{A nontrivial footprint of standard cosmology in
  the future observations of low-frequency gravitational waves},''
  \href{http://arxiv.org/abs/1902.04550}{{\ttfamily arXiv:1902.04550
  [astro-ph.CO]}}.

\end{thebibliography}\endgroup

\end{document}